\documentclass[pra,twocolumn,titlepage,nofootinbib,amsmath]{revtex4}%
\usepackage{amsmath}
\usepackage{amsfonts}
\usepackage{amssymb}
\usepackage{graphicx}%
\newcommand{\rightarrowshort}{{\,\rightarrow\,}}
\begin{document}
\title{Comment on {\em Two-photon approximation in the theory of electron recombination in
hydrogen\/} (D. Solovyev and L. Labzowsky, Phys. Rev. A {\bf 81},
062509 (2010)) }\thanks{An extended version. A short version has
been submitted to PRA \cite{short}.}
\author{S.~G.~Karshenboim}
\email{savely.karshenboim@mpq.mpg.de} \affiliation{Pulkovo
Observatory, St.Petersburg, 196140, Russia}
\affiliation{Max-Planck-Institut f\"ur Quantenoptik, Garching,
85748, Germany}

\author{V.~G.~Ivanov}
\affiliation{Pulkovo Observatory, St.Petersburg, 196140, Russia}

\author{J.~Chluba}
\affiliation{Canadian Institute for Theoretical Astrophysics, Toronto, ON M5S 3H8, Canada}

\newcommand{\eq}[1]{(\ref{#1})}
\begin{abstract}
The results for the total multi-photon decay rates of the $3p$ and
$4s$ levels of hydrogen, presented by D. Solovyev and L. Labzowsky
within the cascade approximation, are revisited. The corrected
results for certain decay channels differ from original ones of
those authors sometimes by order of magnitude.
Some aspects with respect to the cosmological recombination process
are clarified.
\end{abstract}
\maketitle
%


Paper \cite{pra} is devoted to the calculation of the contribution
of various multi-photon decay modes to the lifetime of free
hydrogenic energy levels.
In particular, the multi-photon decays of the $3p$ and $4s$ state
were considered.

Their result for the $3p$ radiative width, which includes
three-photon decay modes (e.g. $3p\rightarrow 2s \rightarrow 1s$)
within the cascade approximation, is given by Eq. (33) of
\cite{pra}:
\[
W^{\rm total}_{3p-1s}=W^{(1\gamma)}_{3p-1s}+\frac34\,
W^{(2\gamma)}_{3p-2p}+\frac34\,
\frac{W^{(1\gamma)}_{3p-2s}}{\Gamma_{3p}}\,W^{(2\gamma)}_{2s-1s} \;.
\]
Here $W_X$ is the probability of the related decay channel $X$, and
$\Gamma_{A}$ is the total radiative width (i.e. the total decay
probability) of state $A$. Clearly, the radiative width, being
calculated for a free atom, satisfies the condition
\[
\Gamma_{A}=W^{\rm total}_{A-1s}\;,
\]
since any excited states should eventually decay into the ground
state after emitting an appropriate number of photons.

Later on, equation (47) of \cite{pra} presents their result for the
$4s$ radiative width, which includes four-photon modes (e.g.,
$4s\rightarrow 3p \rightarrow 2s \rightarrow 1s$) in the cascade
approximation:
\begin{eqnarray}
W^{\rm total}_{4s-1s}&=&W^{(2\gamma)}_{4s-1s}+\frac32\,
\frac{W^{(1\gamma)}_{3s-2p}}{\Gamma_{3s}}\,W^{(2\gamma)}_{4s-3s}\nonumber\\
&&+\frac32\,
\frac{W^{(1\gamma)}_{4s-3p}}{\Gamma_{4s}}\,W^{(2\gamma)}_{3p-2p}
\nonumber\\
&&\quad
+\frac32\, \frac{W^{(1\gamma)}_{4s-3p}}{\Gamma_{4s}}\,
\frac{W^{(1\gamma)}_{3p-2s}}{\Gamma_{3p}}\,W^{(2\gamma)}_{2s-1s}\nonumber\;.
\end{eqnarray}

Although expression (33) of \cite{pra} has the correct order of
magnitude, i.e., $\propto \alpha(Z\alpha)^4m_ec^2/\hbar$, we argue
below that its numerical value is incorrect.
Furthermore, the result given by Eq.~(47) of \cite{pra} is even off by order of magnitude.
The problem is that a conceptual mistake occurred in the calculation
of the cascade terms involving three or four photons.

The appropriate results for $3p$ and $4s$ are well-known within the
cascade approximation\footnote{The cascade approximation for the
dynamics of the decay implies a resonance approximation for the
calculation of the related quantum-mechanical expressions. The
description of various atomic-state decays resulting from the
resonance approximation can be found in standard textbooks.}. The
results for all the decay channels, which contribute in order
$\alpha(Z\alpha)^4m_ec^2/\hbar$, are summarized in Table~\ref{T:3p}
(for the $3p$ state) and in Table~\ref{T:4s} (for the $4s$ state).

\begin{table}[phtb]
\begin{tabular}{l|c|c}
\hline
Channel & ~~Partial width~~ & ~~Partial width in \cite{pra}~~  \\
 \hline
$1\gamma: 3p\rightarrowshort 1s$&$W^{(1\gamma)}_{3p-1s}$&$W^{(1\gamma)}_{3p-1s}$\\
$3\gamma: 3p \rightarrowshort 2s \rightarrowshort 1s$&$W^{(1\gamma)}_{3p-2s}$&$\frac34\,
\frac{W^{(1\gamma)}_{3p-2s}}{\Gamma_{3p}}\,W^{(2\gamma)}_{2s-1s}$\\
\hline
\end{tabular}
\caption{The $3p$ decay channels and their partial width to order
$\alpha(Z\alpha)^4m_ec^2/\hbar$\label{T:3p}}
\end{table}

\begin{table}[phtb]
\begin{tabular}{l|c|c}
\hline
Channel & ~~Partial width~~ & ~~Partial width in \cite{pra}~~  \\
 \hline
$2\gamma: 4s \rightarrowshort 3p \rightarrowshort 1s$&$W^{(1\gamma)}_{4s-3p}\frac{W^{(1\gamma)}_{3p-1s}}{\Gamma_{3p}}$&not specified\\
$2\gamma: 4s \rightarrowshort 2p \rightarrowshort 1s$&$W^{(1\gamma)}_{4s-2p}$&not specified\\
$4\gamma:
4s \rightarrowshort 3p \rightarrowshort 2s \rightarrowshort 1s$&$W^{(1\gamma)}_{4s-3p}\frac{W^{(1\gamma)}_{3p-2s}}{\Gamma_{3p}}$&$
\frac32\, \frac{W^{(1\gamma)}_{4s-3p}}{\Gamma_{4s}}\,
\frac{W^{(1\gamma)}_{3p-2s}}{\Gamma_{3p}}\,W^{(2\gamma)}_{2s-1s}$
\\
\hline
\end{tabular}
\caption{The $4s$ decay channels and their partial width to order
$\alpha(Z\alpha)^4m_ec^2/\hbar$. The $2\gamma$ modes are not
specified in more detail by the authors of \cite{pra}. However, a
related comment in \cite{pra} indicates that some conceptual
differences with our understanding of these channels exist (see
below).\label{T:4s}}
\end{table}

We note that the expressions Eqs.~(29) and (38) of \cite{pra}
introduce the total width of the $3p$ and $4s$ state, respectively.
In the cascade approximation, which is sufficient for calculation of
the leading order contributions and which is supposedly applied in
\cite{pra}, the width of any excited state (except for the $2s$
state) is the sum over $E1$ one-photon decays to all appropriate
lower levels. This value is presented in various textbooks and
summarized in the tables above. Apparently, once the state under
question decays into lower-lying excited states, any further
development due to a subsequent decay of those levels does not
change the width of the initial state, a conceptual aspect that is
different in the analysis of \cite{pra}.

For the $3p$ state there are only two dominant channels, namely a
$1\gamma$ decay ($3p\to1s$) and a $3\gamma$ decay ($3p\to2s\to1s$).
The probability of the second channel, which involves three photons,
is indeed the same as a naive $E1$ $1\gamma$ probability of the
$3p\to2s$ decay, because for a free-atom case 100\% of the atoms in
the $2s$ state decay afterward into the $1s$ state with emission of
two photons. All other channels and any corrections beyond the
cascade approximation are of higher order in $(Z\alpha)$ and thus
can be neglected.
This implies that in particular the last term in Eq.~(33) of
\cite{pra} is incorrect, since it suggests  that the total width of
the $3p$ state is affected by the subsequent decay of the $2s$ state
via two photons.

Technically, the difference originates from the regularization in
Eq.~(29) of  \cite{pra}. Any cascade decay, calculated by means of
Schr\"odinger's equation with a Hermitian quantum-mechanical
Hamiltonian, leads to an expression with a denominator (or few
denominators), value of which vanishes when the photon frequency is
at resonance. The regularization should involve the non-zero width
of the resonant intermediate state (states) as a regulator
(regulators), as e.g. discussed in \cite{letaj}. However, neither
the width of the initial state (as is done in \cite{pra}) nor of the
final state should be introduced. Once we substitute $\Gamma_{2s}$
for $\Gamma_{3p}$ in the denominator, the third term becomes of
correct order. Still it has an incorrect coefficient of (3/4), which
should be replaced by unity.

The second term in Eq.~(33) describes the $3p\to2p\to1s$ channel and
obviously its width should be equal to the width of the $3p\to2p$
decay which appears in $2\gamma$ approximation and is of order
$\alpha^2(Z\alpha)^6m_ec^2/\hbar$. Apparently, the coefficient 3/4
in (33) is again incorrect and should be replaced by unity.
However, although it is clear that a contribution of order
$\alpha^2(Z\alpha)^6m_ec^2/\hbar$ may be of interest for the
differential probabilities of the decay process, it should be
neglected in the total width, since many other corrections of this
order (or even some larger contributions) are not accounted for (see
\cite{letaj,corr} for more detailed discussion).

In the case of the $4s$ state there are three basic modes. Two modes
are for $2\gamma$ decay, namely, $4s\to np\to1s, (n=2,3)$, and one
modes involve four photons ($4s\to 3p\to2s\to1s$). The probability
of the $4s\to 2p\to1s$ mode is the same as the $E1$ $1\gamma$
probability for the $4s\to 2p$, while the sum of decay widths for
the two other channels, namely, of $4s\to 3p\to1s$ and $4s\to
3p\to2s\to1s$, should reproduce the $E1$ $1\gamma$ width of $4s\to
3p$ decay, since the $3p$ state decays into $1s$ either directly or
via the intermediate $2s$ level. The branching ratio for
the $3p$ modes are
\[
Br(3p-ns)=\frac{W^{(1\gamma)}_{3p-ns}}{W^{(1\gamma)}_{3p-1s}+W^{(1\gamma)}_{3p-2s}}\;,
\]
where $n=1,2$. Since all the $2s$ states eventually decay into the
ground state the probability of $3p\to2s\to1s$ decay is equal to the
well-known probability of $3p\to2s$ decay.

For the $4s$ state the leading $4\gamma$ term should read (see
Table~\ref{T:4s})
\[
W^{(1\gamma)}_{4s-3p}\frac{W^{(1\gamma)}_{3p-2s}}{\Gamma_{3p}}\,,
\]
which is to be compared with the last term in Eq.~(47) of
\cite{pra}.
The problem again comes from an incorrect regularization.
Regularizing Eq. (38) properly, one can obtain the correct result
after omitting the pre-factor 3/2 (which should be unity). This
$4\gamma$ term is of the order of the
$\alpha(Z\alpha)^4m_ec^2/\hbar$.

The $2\gamma$ contributions (see Table~\ref{T:4s}) are also of the
same order of magnitude.
The first term in Eq.~(47) of \cite{pra} is intended to take these
contributions into account, but is not specified any further by the
authors of \cite{pra}. The correct $2\gamma$ result for the total
$4s$ width should include cascade contributions, however, a comment
after Eq. (48) of \cite{pra} refers to a numerical value of
12$\;$s$^{-1}$, which seems more consistent\footnote{The value
12$\;$s$^{-1}$ was quoted from Table~1 of  \cite{Chluba2008}.
However, this value was computed using the part of the two-photon
profile that only includes transitions to virtual intermediate
states, which in \cite{Chluba2008} was defined as `non-resonant'
contribution. In \cite{Chluba2008} this definition was merely
nomenclature, and turned out to be convenient in the computation of
the total matrix elements.
But as explained in Sect.~4.3 and Sect.~5 of  \cite{Chluba2008}
because of interference with the resonant contributions to the total
transition matrix element this value {\it should not\/} be
interpreted as two-photon correction to the $4s \to 1s$ transition
rate.}
with a certain tail contribution beyond the
cascade term obviously being of a higher order than
$\alpha(Z\alpha)^4m_ec^2/\hbar$.

The second and the third terms in Eq. (47) are of order
$\alpha^2(Z\alpha)^6m_ec^2/\hbar$ and, similar to our consideration
of the $3p\to2p$ channel, these terms may be in principle of
interest for differential width, but should be neglected in the
total width. Furthermore, the numerical coefficients 3/2 should be
replaced with unity. In addition, for the third term the $4s$ width
used as a regulator in Eqs.~(29) and (38) of \cite{pra} should be
replaced with the $2s$ width.


In general, the cascade approximation cannot help to take into
account `real' multi-photon decay modes. The integral cascade width
is completely determined by the first decay in the chain and does
not involve any information on further subsequent decays.
Calculations of such effects within the cascade approximation was
one of the purposes of \cite{pra}.

There are also problems outside of main consideration of \cite{pra},
which, however, are important for the interpretation of the results.
As we can see, the regularization of quantum-mechanical expressions
(29) and (38) plays a crucial role in calculations. Paper \cite{pra}
is devoted to a free hydrogen atom, but it was motivated by study of
cosmic recombination of hydrogen, which occurred some 380\,000 years
after the big bang, when the Universe had cooled to a temperature of
about $\sim 3000\,$K.
During cosmological recombination, the atoms existed within
an intense bath of the cosmic blackbody radiation.
Under these conditions, the cascade chains should not only include
{\it spontaneous decays\/}, but also {\it excitations\/} and {\it
induced decays\/}, mediated by the cosmic radiation background. This
can change the total width of the $3p$ state by $\sim 1\%$ (see
\cite{Chluba2010} for more details).
Thus the decay width, used as a regulator, should include effects
beyond the free-atom approximation. Notably, in the case of the
cosmic recombination the $2s$ and $1s$ states receive a width
induced by the blackbody CMB radiation \cite{Chluba2006}.

Next, we have to check whether the width of the initial and final
states are important for the consideration.
Any partial width should be calculated without introducing the width
of the initial state into any denominator of expressions similar to
Eqs.~(29) and (38) of \cite{pra}. Nevertheless, the width of initial
state may appear in certain expressions, however for a different
reason.
It enters through branching ratios (relative probabilities) for the
transitions of interest, which needs to deal with combinations such
as
\[
\frac{W_{3p-2s}}{W_{3p-2s}+W_{3p-1s}}=\frac{W_{3p-2s}}{\Gamma_{3p}}\;.
\]
This expression determines the portion of $3p$ states that decay
with the emission of three photons (assuming a free decay, where the
dominant mode of $2s$ decay is a two photon process).
This has to be used in the denominator as total width of the initial
state (which in our example is the $3p$ state).

The initial- and final-state widths are even more important in
another way. Once we want to consider the dynamics beyond the
cascade approximation or wish to derive the cascade approximation
from rigorous quantum-mechanical expressions, we have to start with
a certain expression similar to Eqs.~(29) and (38) of \cite{pra}.
However, we have to start such an evaluation with `quasi-stable'
initial and final levels. The conditions
\[
\frac{\Gamma_{\rm initial}}{\Gamma_{\rm intermediate}} \ll 1
\]
and
\[
\frac{\Gamma_{\rm final}}{\Gamma_{\rm intermediate}} \ll 1
\]
are necessary to validate such an approach.

The cascade approximation means that all the levels are created,
propagate and decay in a factorized way. E.g. the lifetime of an
`initial' or `intermediate' state, and details of their decay do not
depend on a way they have been created.
Meantime, the expressions such as Eqs.~(29) and (38) pretend to go
beyond such a factorized description. However, the very formulation
of the problem, such as a decay of the $3p$ or $4s$ state, means
that we already partly consider a cascade approximation, because the
very existence of those states as initial states means that we
ignore details of their creation.

As is well-known, off-resonance corrections are larger for broad
levels than for narrow ones, and it is reasonable to consider
several most narrow levels in a pure resonance approximation. The
very consideration of a decay of a certain state into a set of final
states within an approach given by Eqs.~(29) and (33) of \cite{pra}
is meaningful only if the initial state together with all possible
final states of the decay chains is more narrow than any
intermediate state, which means ${\Gamma_{\rm init}}/{\Gamma_{\rm
interm}} \ll 1$ and ${\Gamma_{\rm fin}}/{\Gamma_{\rm interm}} \ll
1$.

For example, if we consider a frequency distribution of emission
lines, the line width of a particular resonance photon is determined
by the width of both initial and final states. That means that in a
chain of transitions we need to take into account both widths, and
thus both the states should be treated as metastable (unstable) for
the same reason. We cannot really consider any of them as an
`initial' or `final' state of a cascade chain. We need to introduce
creation of the initial state and decay of the final state, unless
one of them lives much longer than the other and its width can be
neglected.

For the consideration of the $3p-1s$ three-photon decay with a
resonance at $2s$ the width of resonance $3p-2s$ photon is
determined by the ambiguity in the very formulation of the problem
of decay of $3p$ state, while the width of sum of two frequencies of
the $2s-1s$ resonance is determined by the $2s$ width. The
uncertainty in energy of the initial state is more important than
the width of the $2s$ state. That invalidates the very consideration
of decay of any state (such as $3p$ or $4s$) into the $1s$ state via
the $2s$ state. In principle, such a consideration should consider
the $2s$ state as a stable one.

However, if we are to eventually arrive at a pure resonance
approximation it is not important in which order we `break' the
chain and which levels we already consider in the cascade
approximation. Finally, all accessible intermediate states become
resonances. Since paper \cite{pra} presumes to derive the results in
a pure resonance approximation, the formulation of the problem of
the decay of the $3p$ and $4s$ state is not quite correct, but
should eventually produce correct results. That is because of the
fact that there are two kinds of parameters. One is for the ratio of
a width and a characteristic frequency, and the other is ratios of
different widths. The latter are important to partially consider
dynamics beyond the cascade descriptions. The former are always
small by a factor $\alpha(Z\alpha)^2$ or less and they are
sufficient to derive the cascade results.

Consideration of any modes beyond the leading contributions, which
are with one-photon decays of any initial state to lower states, are
meaningless for the integral line width, but may be important for a
differential width as explained in \cite{letaj}. Indeed, there is no
real separation between tail of the `resonance' terms and
`off-resonance modes' and interference terms and for the
differential effects one has to deal with a complete width.

This aspect of the problem is also important for recent computations
of the cosmological recombination process \cite{Chluba2010, Haimoud2011}, where
deviations of the differential cross-sections from the normal
Lorentzian profile \citep{Chluba2008, Hirata2008} are accounted for, in both
{two-photon decay channels} (e.g., $3d\rightarrow2p\rightarrow
1s$) and {Raman-events} (e.g. $2s\rightarrow 3p \rightarrow
1s$). No explicit separation in cascade or off resonance
contributions is made, but the total interaction of atoms with the
ambient cosmic radiation background, including {photon
production}, {photon absorption}, and {photon scattering},
are taken into account consistently.

The work was supported in part by DFG (under grant \# GZ HA
1457/7-1) and RFBR (under grant \# 11-02-91343). Discussions with
R.A. Sunyaev are gratefully acknowledged.

\end{document}